\documentstyle[12pt]{article}
\title{{\bf Symmetry and interactivity in programming}}
\date{October 2001} 
\author{P.-L. Curien}

\begin{document}
\maketitle

\newtheorem{theoreme}{Th\'eor\`eme}[section]
\newtheorem{definition}[theoreme]{D\'efinition}
\newtheorem{lemme}[theoreme]{Lemme}
\newtheorem{corollaire}[theoreme]{Corollaire}
\newtheorem{proposition}[theoreme]{Proposition}
\newtheorem{exemple}[theoreme]{Exemple}
\newtheorem{exercice}[theoreme]{Exercice}
\newtheorem{hardexercise}[theoreme]{Exercise${}^*$}
\newtheorem{remarque}[theoreme]{Remarque}

\newenvironment{branch}{\left\{\begin{array}{l}}{\end{array}\right.}

\newcommand{\qedm}{\mbox{}\\[-2.5em] \mbox{}\hfill\rule{6pt}{6pt}} 
\newcommand{\Preuve}{\noindent {\sc Preuve}. }
\newcommand{\Preuvehint}{\noindent {\sc Preuve (indication)}. }
\newcommand{\Proofitem}[1]{\medskip \noindent $#1\;$}
\newcommand{\Proofitemf}[1]{ $#1\;$}

\newcommand{\guil}[1]{``#1"}
\newcommand{\lbd}{\lambda}
\newcommand{\mand}{\mbox{ and }}
\newcommand{\mor}{\mbox{ or }}
\newcommand{\nin}{\not\in}
\newcommand{\inc}{\subseteq}
\newcommand{\set}[1]{\{#1\}}
\newcommand{\setc}[2]{\set{#1 \mid #2}}
\newcommand{\faux}{{\bf F}}
\newcommand{\et}{\wedge}
\newcommand{\ou}{\vee}
\newcommand{\vide}{\emptyset}
\newcommand{\si}{\Leftarrow}
\renewcommand{\iff}{\Leftrightarrow}
\newcommand{\qqs}[2]{\forall\, #1\;\: #2}
\newcommand{\xst}[2]{\exists\, #1\;\: #2}
\newcommand{\nxst}[2]{\not\!\exists\, #1\;\: #2}
\newcommand \seql[3]{\raisebox{3ex}{$\mbox{#1}\;\;$} \; \shortstack{$#2$ \\ \mbox{}\\
                    \mbox{}\hrulefill\mbox{}\\ \mbox{}\\ $#3$}}
\newcommand \seq[2]{\shortstack{$#1$ \\ \mbox{}\\
                    \mbox{}\hrulefill\mbox{}\\ \mbox{}\\ $#2$}}

\newcommand{\forces}{\makebox[5mm]{\,$\|\!-$}}%

\newcommand{\fun}{\:\rightarrow\:} 
\newcommand{\funt}{\rightarrow^{\ast}} 
\newcommand{\lfun}[1]{\stackrel{#1}{\longrightarrow}}
\newcommand{\lfunt}[1]{\stackrel{#1}{\longrightarrow}\,\!^{\ast}}
\newcommand{\length}[1]{ | #1 |}
\newcommand{\union}{\cup}		
\newcommand{\inter}{\cap}		
\newcommand{\Union}{\bigcup}		
\newcommand{\Inter}{\bigcap}		
\newcommand{\join}{\vee}		
\newcommand{\JOIN}{\bigvee}		
\newcommand{\meet}{\wedge}		
\newcommand{\MEET}{\bigwedge}		

\newcommand{\inv}[1]{#1^{-1}}
\newcommand{\card}{\sharp}		
\newcommand{\pfun}{\rightharpoonup}  
\newcommand{\ul}[1]{\underline{#1}}	

\newcommand{\id}{{\it id}}
\newcommand{\catch}{{\it catch}\;}
\newcommand{\throw}{{\it throw}\;}
\newcommand{\Alt}{ \mid\!\!\mid  } 
\newcommand{\Sub}[3]{#1[#2\leftarrow #3]}	
\newcommand{\sub}[2]{[#2 / #1]}
\newcommand{\subn}[2]{[#2 /_{\!\circ} #1]}
\newcommand{\BO}[4]{#1\,,#2\stackrel{#3}{\Longrightarrow} #4}
\newcommand{\bt}[1]{{\it BT}(#1)}
\newcommand{\lelong}[2]{#1\!\mid_{#2}}	
\newcommand{\dl}{[\![} 			
\newcommand{\dr}{]\!]} 			
\newcommand{\newv}{{\it new}\:}
\newcommand{\deref}{{\it deref}\:}
\newcommand{\nat}{{\bf int}}
\newcommand{\varimp}{{\bf var}\:}
\newcommand{\comm}{{\bf comm}}
\newcommand{\world}{{\bf W}}
\newcommand{\comp}{\circ}
\newcommand{\run}{{\it run}}
\newcommand{\done}{{\it done}}
\newcommand{\readv}{{\it read}}
\newcommand{\writev}[1]{{\it write}(#1)}
\newcommand{\OK}{{\it OK}}

\begin{abstract}
We recall some of the early occurrences of the notions of interactivity and symmetry in the operational and denotational semantics of programming languages. We suggest some connections with ludics.
\end{abstract}

Proof theory and programming are deeply intertwinned research areas. This short paper will address and compare ideas from both sides, so it seems appropriate to begin by recalling what this general interplay is about, before we enter into the more specific topic we have in mind here, namely symmetry and interactivity of computation.

\medskip
First, a disclaimer is in order: the encounter of logic and programming is by far not limited to what is commonly called logic programming. This expression refers to a programming style in which computation is proof search.  
I am speaking here of the larger match between two kinds of wide-range activities: 

\begin{itemize}
\item writing programs, or any piece of structured software, 
proving their correctness with respect to a specification, designing a language, implementing it via a compiler into another programming language, which may be itself amenable to the same kind of analysis, etc...;
\item formalizing proofs -- interesting ones rely on various lemmas, sublemmas, etc... --, checking them, or searching them, designing a logical system, encoding  a logic into another, etc... .
\end{itemize}

This correspondence, or lexicon, has been working extremely well in the past two decades, for the mutual benefit of the two disciplines. It has underpinned the development of logic programming and of functional programming, based on the paradigms of proof-search and proof-transformation, respectively, and also of proof assistants such as the system Coq based on the Calculus of Constructions of Coquand and Huet.

The lexicon is not just a term-to-term correspondence. It also concerns the {\em internal structure} -- that is, the underlying engine -- deeply. On one hand, the cut-elimination procedure transforms an \guil{implicit} proof with lemmas (called {\em cuts} in logic) into an entirely explicit proof, and as a benefit allows for an analysis of the resulting cut-free proof that enjoys the subformula property (every formula appearing in the proof appears also in the conclusion). On the other hand, the normalization procedure, or more generally an operational semantics for a programming language, describes formally the computation steps of the programs. It turns out that cut-elimination in logic, and normalization in formal semantics of programming languages match closely. It has been remarked by Howard, who pointed out in \cite{How} that there exists a one-to-one correspondence between proofs in natural deduction style and (typed) 
$\lambda$-terms: this is the celebrated {\em Curry-Howard} isomorphism.
More recently, Herbelin \cite{Her95} 
noticed that $\lbd$-caclulus could be seen as matching Gentzen's {\em sequent calculus} closely and directly, which was a door open to exhibiting more symmetries in syntax \cite{CH00}.

A fundamental new theoretical input of this period was Girard's linear logic \cite{Gir87}, which entailed a thorough revisiting of proof theory, but also of the semantics of programming languages.
A keyword behind linear logic is \guil{resource consciousness}, which in turn may have a declination in time and in space. Linear logic focused on time, while the latest body of work of Girard -- ludics --
is also concerned with space. 
The main source of time complexity in the cut-elimination procedure comes from repeated uses of assumptions, which corresponds to duplication of  computations.
In linear logic, only formulas with an explicit modality (noted !) can be freely reused.
Space consciousness, on the other hand, amounts to taking into account memory management.
Time and space: think of your favourite home dictionary: it may have been used so many times that it is now in a very poor condition; or you may decide to put it away in your bookcase, but you discover that there is no room left for it on the shelves...

\medskip
Important transfers inspired by the above lexicon include a
revisiting of proof search in logic programming influenced by linear logic \cite{AP-Foc}, which in turn has been influential on the design principles of ludics, or the extension of the Curry/Howard correspondence from pure effect-free programs to ones with effects such as the raising of exceptions \cite{Grif}. The so-called continuation-passing-style translations developed by computer scientists to capture the meaning of such effects happen to match precisely so-called double-negation translations
(from classical to intuitionistic logic)  designed by logicians much earlier.

\subsection*{Some dual pairs}
It is remarkable that largely independent developments have converged to the recognition of the importance of
{\em symmetries} in both worlds, and then of the paradigm of {\em computation-as-interaction}.
Let me emphasize this from  the computer science standpoint first.
Here are a few often found  (and related) complementary pairs:
\begin{itemize}
\item a memory {\em cell} or location or register versus its actual contents or {\em value}; in object-oriented style, the record field names versus their values, the method names versus their actual definition;
\item {\em input} and {\em output};
\item {\em sending} and {\em receiving} messages;
\item a {\em program} and its {\em context} (the libraries of your program environment -- or the larger program of which the program under focus is a subpart, or a module); the programmer and the computer; two programs that call each other.
\end{itemize}

One may also add vaguer, but helpful complementary pairs, taken from \guil{human life}: questions and answers, attack and defence, or -- more down-to-earth! --
male and female plugs. Such a \guil{computer/man} style is often helpful informally, as in the sentences:
\guil{the programmer interacts with the computer through the top-level loop, typing commands and waiting for the prompts (or error messages)}, or: \guil{a program $A$ computing a function $f(a,b,c)$ of three arguments of boolean type  has used only the first two arguments, and more precisely has first explored the second argument; as this argument was itself the output of another program $B$, program $A$ passed the control to program $B$, waiting for $B$'s output or response; when $B$ has finished, it passes the control back to $A$, which may proceed for a while and
then wait for the output of a program $C$ computing the first argument, etc...}. In the programming language community, the latter mechanism is known as that of {\em coroutines}. It is interactive in nature: $A$, $B$, and $C$ exchange questions and answers, in an incremental way, i.e., when $B$ is resumed, it starts from the point of control that it had reached at the moment it had last returned an answer to $A$.

This model of computation is so important and simple that it deserves an illustration by a simple example. Suppose you have a tree, whoses leaves are labelled by numbers (the internal nodes do not have labels, or they don't matter). One whishes, say, to add all the numbers at the leaves. A simple recursive definition will do, but it embodies a commitment to inorder traversal of the tree. The coroutine view allows us to parameterize over the choice of the traversal algorithm (program $A$), and over the computation we want to do with the labels of the leaves (program $B$). For example, one may take for program $A$ a post-order traversal procedure, and for program $B$ a simple procedure that adds its argument to 
the contents of a fixed register and returns the new value of the register.

\medskip
Interaction is also central in models of concurrent computation such as Milner's CCS or the $\pi$-calculus \cite{Milpi}. There, 
programs and agents may be active simultaneously but have to synchronize 
at appropriate points of their execution. The synchronization may involve the transmission of a value, one process being the sender and the other the receiver. 
The logical counterpart of process calculi is yet to be found. Some believed to find it in linear logic. Ludics may offer  better chances, because of the importance of naming in these calculi (names of channels, of intermediate data or computations, etc...), and clearly names have to do with memory locations.

\medskip
In logic, the related pairs are the following:
\begin{itemize}
\item {\em Hypotheses} and {\em conclusions}: Gentzen's sequent calculus manipulates formal {\em sequents} of the form $\Gamma\vdash\Delta$, where $\Gamma$ and $\Delta$ are sequences of formulas, to  be interpreted as: \guil{the conjunction of the formulas in $\Gamma$ entails the disjunction of the formulas in $\Delta$}. In sequent calculus, each connective comes with two introduction rules, one on the left of $\vdash$, the other on the right.
Symmetry goes with the involutive character of negation, which is lost in intuitionistic logic. 
Thus sequent calculus (for classical logic) enhances symmetries, but cut-elimination, besides being inherently non-deterministic (whence the restrictions to intuitionistic subsystems), is rather clumsy (due to boring \guil{commutative conversions}). The novelty of linear logic was to offer both symmetry {\em and} a satisfactory computational behaviour (confluence and normalization).
  
\item Proofs can be given a \guil{dialogue-game} interpretation, and are then called {\em strategies}. This was first observed
by Lorenzen and his coworkers in the early fifties (see e.g. \cite{Fel86}). Under this interpretation, a formula is checked for correctness through a dialogue between an opponent who doubts some formulas, and the player who justifies his proof step-by-step by exhibiting the rules he has used, in a top-bottom way. The dialogue begins as follows: the opponent challenges the player to justify the conclusion $A$. The player then exhibits
the inference rule he has used last, which has, say, $A_1,\ldots,A_n$ as antecedents. Then the opponent picks one of the formulas, say $A_i$, and again challenges the player to justify how he has reached  this intermediate conclusion, etc...  Note that in this approach, each dialogue is far from amounting to a full proof: it is the collection of all such dialogues or experiments which characterizes the proof. This is very much related to 
the notion of observation at the basis of the study of the equivalence of programs and processes in computer science.  Unfortunately, the dialogue games' school remained at a static, descriptive level. The full flavour comes when cut-elimination is interpreted as a play \cite{Coq} (see also below).
\end{itemize}

\subsection*{Computation as interaction}

From the computer science perspective, the history of the computation-as-interaction paradigm is inseparable from the study of sequentiality. Vuillemin, and Milner have given the first denotational definitions of a sequential function, which were later generalized by Kahn and Plotkin to the framework of concrete data structures (see e.g. \cite{AmaCu}). 
Intuitively, a sequential function is one for which a sequential schedule can be given. The best way to grasp it more exactly is to think of a sequential function as one of the programs in a system of coroutines. In the example above, program $A$ could be $$\mbox{if } b= \mbox{true then if }a=\mbox{true then true}$$

\noindent
which is a partial specification of the logical {\it and} function of the first two arguments. The function $f(a,b,c)$ associated with the program is sequential because it can be scheduled: first compute $b$, then $a$. Both arguments $b$ and $a$ are needed. 
Note that the following is another sequential schedule $A'$ for the same function:
$$\mbox{if } a= \mbox{true then if }b=\mbox{true then true}$$

\noindent
Some functions in denotational semantics cannot be scheduled in this way, the simplest example being Plotkin's parallel-or function ${\it por}$, which is such 
that ${\it por}(\mbox{true},\bot)=\mbox{true}$ and 
${\it por}(\bot,\mbox{true})=\mbox{true}$. In these equations, $\bot$ is a symbol for an undefined value. (One endows ${\bf B}=\set{\bot,\mbox{true},\mbox{false}}$ with the partial order defined by $\bot\leq\mbox{true}$ and $\bot\leq\mbox{false}$.) A program implementing ${\it por}$ should be able to output the value true as soon as either $a$ or $b$ is true, and supposing that $a$ and $b$ are the outputs of two programs $P$ and $Q$ it would necessitate computing $P$ and $Q$ in parallel.
A more subtle example was given by Berry and is known as Gustave's function, or Berry-Kleene function (as Kleene had encountered a variant of this function  too):
$$\begin{array}{l}
BK(\mbox{true},\mbox{false},\bot)=\mbox{true}\\
BK(\mbox{false},\bot,\mbox{true})=\mbox{true}\\
BK(\bot,\mbox{true},\mbox{false})=\mbox{true}
\end{array}$$

\noindent
This function too cannot be scheduled sequentially.
In the mid-seventies, Berry worked on the modelling of the notion of sequential computation and isolated the important intermediate notion of stability (which later led Girard, independently, to linear logic). The functions $por$ and $BK$  are typical examples of a continuous and non stable function, and of a stable and non sequential function, respectively. Berry also  had the insight that to model the sequential computational behaviours, and only them, one should move from a traditional framework of appropriate ordered sets (domains in the jargon of denotational semantics) and {\em functions} to a setting retaining more than the ordinary input-output behaviour of programs. For example,
the two above schedules for the logical {\it and} function are given different interpretations in the model.

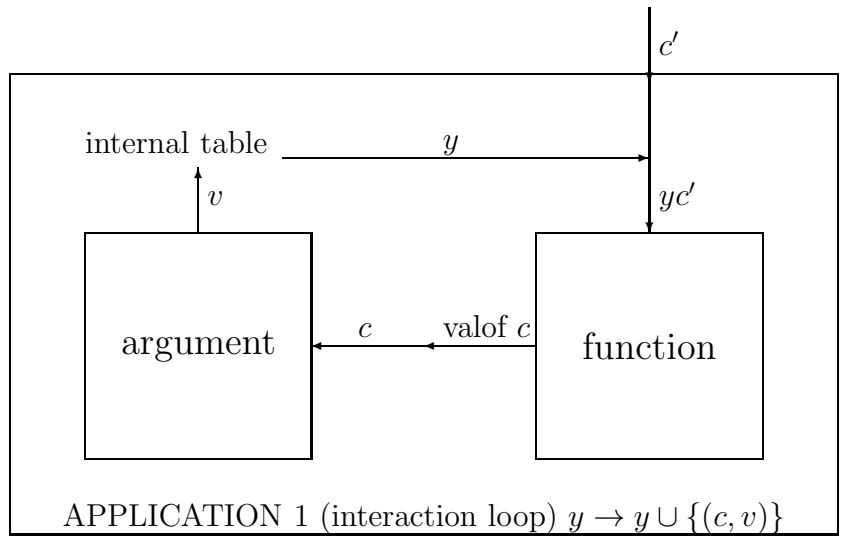
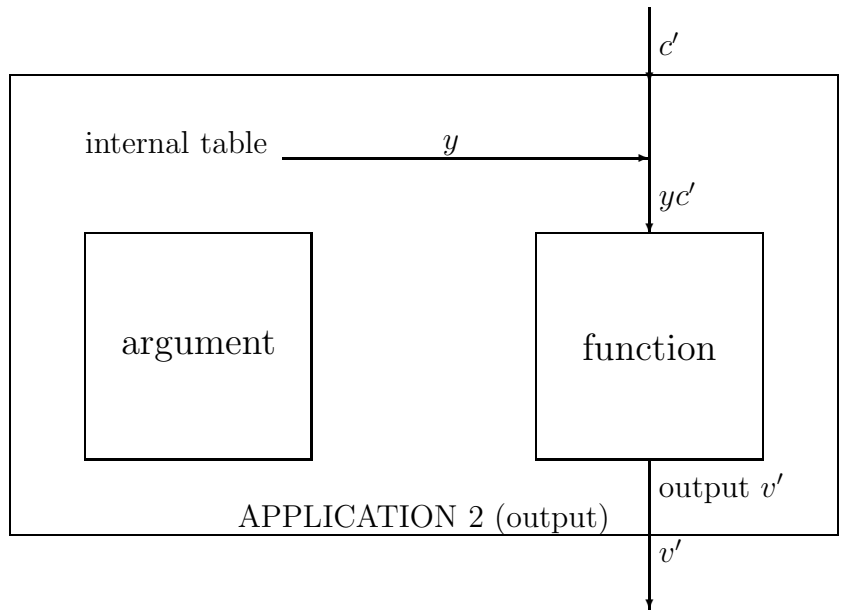
\setlength{\unitlength}{0.5cm}
\begin{picture}(30,40)(0, -10)
\put(0,0){\framebox(22,12)[bc]{APPLICATION 2 (output)}}
\put(2,10.15){\makebox{internal table}}
\put(2,2){\framebox(6,6){{\large argument}}}
\put(14,2){\framebox(6,6){{\large function}}}
\put(17,14){\vector(0,-1){2}}
\put(17.25,12.75){\mbox{$c'$}}
\put(17,12){\vector(0,-1){4}}
\put(17.25,8.75){\mbox{$yc'$}}
\put(7.25,10){\vector(1,0){9.75}}
\put(11.5,10.25){\mbox{$y$}}
\put(17,2){\vector(0,-1){4}}
\put(17.25,1){\mbox{output $v'$}}
\put(17.25,-0.75){\mbox{$v'$}}

\put(2,28.15){\makebox{internal table}}
\put(0,18){\framebox(22,12)[bc]{APPLICATION 1 (interaction loop) $y\rightarrow y\union\set{(c,v)}$}}
\put(2,20){\framebox(6,6){{\large argument}}}
\put(14,20){\framebox(6,6){{\large function}}}
\put(5,26){\vector(0,1){1.75}}
\put(5.25,26.75){\mbox{$v$}}
\put(11,23){\vector(-1,0){3}}
\put(9.25,23.25){\makebox{$c$}}
\put(14,23){\vector(-1,0){3}}
\put(11.5,23.25){\mbox{valof $c$}}
\put(17,32){\vector(0,-1){2}}
\put(17.25,30.75){\mbox{$c'$}}
\put(17,30){\vector(0,-1){4}}
\put(17.25,26.75){\mbox{$yc'$}}
\put(7.25,28){\vector(1,0){9.75}}
\put(11.5,28.25){\mbox{$y$}}
\end{picture}

The insight resulted in the  model of sequential algorithms \cite{BerryCurien82}, which I presented in 1978 in a Spring School on $\lbd$-calculus in La Ch\^atre. In this model, morphisms are not functions but pairs of  a function and a computation strategy for it, that specifies a schedule of interaction of the function with its argument. 

The model was then turned into syntax, and a programming language called CDS was developped \cite{BerryCurien85}. The operational semantics of the language, which I presented in 1982 in a joint French-US workshop held in Fontaine\-bleau, was -- as far as I know -- the first appearance of the notion of function application (or function composition) as a dialogue, from an operational point of view. We shall briefly explain how it works. A concrete data structure  has a collection $C$ of cells and a collection $V$ of values.  Datas $x$ are represented as sets of pairs $(c,v)$ such that $c\in C$ and $v\in V$. Hence a data is made of elementary bricks which consist of a cell $c$ filled with some value $v$. For example, the triplet $(\mbox{true},\mbox{false},\bot)$ may be represented as 
$\set{(a,\mbox{true}),(b,\mbox{false})}$: there are three cells corresponding to the three coordinates $a,b,c$, and we spell out that $a$ holds the value true, that $b$ holds the value false, and that $c$ is not filled. 

In CDS, computation proceeds in a lazy, stream-like way. If an expression $a$ of the language, of any type, is entered for evaluation, the interpreter of the language prompts the user for a request $c$ and returns the value $v$ of this cell in (the meaning of) $a$, and then prompst the user for a new request $c_1$, etc...

 The cells of a function type have the form $xc'$, and the values have the form  \guil{$\mbox{valof }c$} and \guil{$\mbox{output }v'$}. If $f$ is a sequential algorithm, then 
$(xc',\mbox{output }v')\in f$ (roughly) expresses that $(c',v')\in f(x)$, and
$(xc',\mbox{valof }c)\in f$ expresses that, at input $x$, in order to compute the value of the output cell $c'$, $f$ needs, or waits for, the value of cell $c$. Now we are ready to explain the dynamics of function application, as illustrated by the two figures APPLICATION 1 and APPLICATION 2: $f(x)$, when presented with a request $c'$, consults the \guil{state} $y$ of an internal table, which stands for the part of the input $x$ read so far (initially, the internal table is the empty set). The function is then asked the question $yc'$. If may either answer (cf. above) \guil{$\mbox{valof }c$} or \guil{$\mbox{output }v'$}. In the first case, control is transfered to the argument $x$, because the value of cell $c$ in $x$ is requested. When $c$ answers with some $v$, then the internal table is updated and become $y'=y\union\set{(c,v)}$, allowing for the more informed
question $y'c'$ to be asked. In this way a loop is formed between $f$ and $x$, whose \guil{trace} consists of a sequence $c,v,c_1,v_1,\ldots$ of alternating \guil{opponent and player moves} (following the above dictionary of dual pairs). The loop terminates when $f$ has received enough information from $x$, i.e., when the internal table $y$ has become large enough, so that
$f$ presented with $yc'$ answers with \guil{$\mbox{output }v'$}, which yields $v'$ as answer to the initial request $c'$ to $f(x)$.

To be fair, this interactive sort of semantics was not the intended goal when we started. Our motivation was to provide a \guil{denotational} description of the fully abstract model of PCF. The language PCF is a simply typed $\lambda$-calculus with constants and recursion that encodes all partial recursive functions, and a fully abstract model is one in which two terms $M$ and $N$ receive the same interpretation if and only if they cannot be separated by any observation (cf. 
the dialogue interpretation of proofs, above). An observation on a term $M$ is defined with the help of a context $C$ (a term of base type with a hole): one evaluates $C[M]$, which yields a base constant $c$ or does not terminate. Two programs can be separated if there exists a context $C$ for which $C[M]$ and $C[N]$ do not evaluate in the same way.

Sequential algorithms did provide a fully abstract model, not for PCF, but for extensions of PCF with control primitives, as offered in the language CDS, or in PCF extended with an operation ${\it catch}$ \cite{CCF94}. The model is effective in the sense that observational equivalence classes can be effectively enumerated, and there is even a finite number of them when the base types are themselves finite (such as the Booleans).
Again, this full abstraction result was not the intended goal: we wanted a fully abstract model of PCF for short. But Loader's later result \cite{Loa} settled the question negatively: he showed that the observational equivalence for PCF is not effective. As a matter of fact, the game-theoretic models of PCF given in 1994 by Abramsky, Jagadeesan, and Malacaria (AJM), and by Hyland and Ong (H0) \cite{HO,AJM} offer syntax-free presentations of {\em term-models}, and the fully abstract model of PCF is obtained from them by a rather brutal quotient, called \guil{extensional collapse}, which gives little more information than Milner's original term model construction of the fully abstract model.

With hindsight, the full abstraction problem was a very interesting, but poorly specified, problem. One looked for a \guil{domain theory}-like presentation of the fully abstract model which was known to exist and to be unique \cite{Mil77}.
But what domain-like meant exactly was not really spelled out. In particular, the effectivity criterion came only to light when it was made possible to contrast different sorts of game models. Indeed, the model of sequential algorithms was given a game-theoretic presentation by Lamarche \cite{Lam92}: the main difference with the AJM and HO models lies in the definition of the $!$ connective, which is set-based for sequential algorithms (whence its finitary character) and multiset-based for the AJM and HO models.

But this poorly stated long standing \guil{open problem} did trigger an important amount of works which are often more important for their side effects. The variety of games models created a new era in denotational semantics, and the HO presentation of games led to an important classification of some features of sequential programming languages such as control, or references (see \cite{AMC} for a survey). The whole approach 
received many insights from the developments of linear logic, and in particular of the geometry-of-interaction interpretation of linear logic  
\cite{Gir89,AJ92}. For instance,
Lamarche's decomposition of the function space in the model of sequential algorithms allowed me to give a more symmetric presentation of affine sequential algorithms, as pairs of two functions, one from input data (or  strategies, in game-theoretic terms) to output data, the other from output {\em counter-strategies} to input counter-strategies. In terms of the discussion above, the two functions take care of the
pairs $(xc',\mbox{output }v')$ (input $x$, piece of output $(c',v')$) and $(xc',\mbox{valof }c)$ (\guil{input} $c'$, \guil{output} $c$), respectively \cite{Symseq}.

Last but not least, we discovered later that Kleene had the same experience of a need to record internal information in addition to the plain underlying functions, and as a matter of fact he  essentially built the Berry-Curien sequential algorithms at lower order types \cite{Kleene}. In his flourished vocabulary, he
 modelled higher-order recursive computations as \guil{machines} communicating via \guil{oracles}, or \guil{envelopes} that are handed by the sender and opened by the receiver.

\medskip
In the early nineties, Cartright and Felleisen found another presentation of sequential algorithms, as \guil{input-output} functions. This is achieved through the introduction of {\em error values} in the semantics. In the presence of these new values, the computation strategy can be made visible interactively, i.e., is part of the input-output behaviour. Take the above $A$ and $A'$:
$$\begin{array}{l}
\mbox{if } b= \mbox{true then if }a=\mbox{true then true}\\
\mbox{if } a= \mbox{true then if }b=\mbox{true then true}
\end{array}$$

\noindent
We have (forgetting about the dummy argument $c$):
 $A(err,\bot)=\bot$ and $A'(err,\bot)=err$, i.e. 
$A$ and $A'$ are different functions (note that $A$ and $A'$ as functions differ only on data containing errors values). 
The informal explanation of these values is as follows: $A$ needs its second argument, and hence cannot output anything if the second argument is $\bot$;
$A'$ needs its first argument, and if this argument is $err$, this stands for some error that occurred when computing the first argument, which terminates the whole computation, so that the value $err$ is propagated to the top-level (second occurrence of $err$ in $A'(err,\bot)=err$).

It was  a striking discovery for me to realize that Girard has introduced the same notion, on the logical side (dynamic dictionary!), which he calls the Demon. The Demon is placed somewhere in a proof which the player does not want to justify completely, or in a counter-proof or observation which the opponent wants to terminate. Think of a human situation where you have a few questions to ask to someone, and where you want to stop the conversation when you have got 
the answers you wanted to get. In the same way, errors help to terminate a computation that reveals a part of the behaviour of a program: the observation
encoded by the argument (or counter-strategy) $(err,\bot)$ reveals that $A'$ starts by examining its first argument.

\subsection*{Interactive types}

The game-theoretic approach to semantics allows us to achieve a better match between primitive base types and defined data types. We illustrate this point with the Booleans. Recall that
we interpret the boolean type as ${\bf B}=\set{\bot,\mbox{true},\mbox{false}}$, where the order is given by
$\bot\leq\mbox{true}$ and $\bot\leq\mbox{false}$. 
The logicians have shown that Booleans can be defined in second-order logic as
$\qqs{X}{X\rightarrow (X\rightarrow X)}$. It was observed by Laird that there exists a suitable game $o$ for which $o\rightarrow (o\rightarrow o)$ is
isomorphic to {\bf B} \cite{Laird}. To begin, note that this is not true with usual domains: either $X$ consists of just $\bot$, and then $X\rightarrow (X\rightarrow X)$ is also a singleton (the map that sends $\bot$ and $\bot$ to $\bot$), or $X$ has at least one non-bottom element, and then it is easy to see that there will be strictly more than 3 elements in (the interpretation of) 
$X\rightarrow (X\rightarrow X)$. But with games, one may consider a structure $o$ with just one cell -- called \guil{$?$} --, and no value. Then a sequential algorithm viewed as a strategy can only be one of the three following
sequences of moves (we write $o\rightarrow o\rightarrow o$ as
$o_{11}\rightarrow o_{12}\rightarrow o_1$, and we place corresponding subscripts to the unique move of each copy of $o$, thus, e.g., $?_{12}$ refers to a move in $o_{12}$):
$$\begin{array}{l}
\mbox{the empty sequence of moves}\\
?_1 \; ?_{11} \\
?_1\; ?_{12}
\end{array}$$

\noindent
or, with the notation used above: 
$$\set{} \quad \set{(\set{}?_1,\mbox{valof }?_{11})} \quad \set{(\set{}?_1,\mbox{valof }?_{12})}\;.$$

\noindent
There is nothing to output, so the algorithms are mere schedulers here. Now, it is straightforward to match these three algorithms with $\bot$, true, and false, respectively (and this is in agreement with the $\lambda$-calculus encoding of 
$\lbd xy.x$ as true and $\lbd xy.y$ as false).

\medskip

Following Girard's notion of {\em behaviour} in ludics, the interactive approach also allows us to enlarge the notion of \guil{type}. A type is then a
collection of programs that behave the same way, in reaction to a set of experiences or tests, which are nothing else than other programs. Taking images of ordinary life, this makes perfect sense: everyone can find for himself examples of how a change in his environment may have affected her or his behaviour.

Let me give an example of type/behaviour, which may have potential applications. One may take as test set a single \guil{taster} program $A$ of type $((\sigma_{111}\times\sigma_{112}\times\sigma_{113}\rightarrow\sigma_{11})\rightarrow\sigma_1)$
that examines its argument -- an algorithm of three arguments -- and returns an error value if, say, the function says that it needs its $i$th argument. If its argument does something else, like outputing directly a value without looking at its argument, or says it needs its $j$th argument ($j\neq i$), then the taster does not proceed further, and in particular does not  deliver any error or non-error value.
Formally, we write (for $i=2$):
$$A=?_1\;?_{11}\;?_{112}\;{\it err}\;.$$

\noindent
Then it is easy to see that the behaviour $A^\bot$ consisting of all programs {\em orthogonal} to the taster, i.e., such that the interaction with the taster yields \guil{error} (or Demon, cf. above),
is the set of algorithms $B$ that start by examining their second argument, or, equivalently, such that $B(\bot,{\it err},\bot)={\it err}$.
This sort of neededness analysis is extremely useful in practice. Remember that the \guil{Ariane V bug} was caused by an overflow due to a piece of useless data....

\medskip
This notion of type is well-suited to the semantics of subtyping: a subtype of a behaviour $X^\bot$ (where $X$ is a set of tests) is just a behaviour of the form $Y^\bot$ for a larger set of tests, i.e., $X\inc Y$ (and 
$Y^\bot\inc X^\bot$). For example, the type of records with fields \guil{year},
\guil{price}, and \guil{colour} is a subtype of the type of records with the fields
\guil{year} and
\guil{price}. A record of the latter type is interactively recognized by testing the presence of the two fields \guil{year} and
\guil{price}, while a third test for the presence of \guil{colour} is needed to \guil{type-check} the membership to the subtype. 
Moreover, intersection types are just usual intersections: this is what Girard calls the locative point of view, but we must end this exposition somewhere, and refer to \cite{Gir01}.

\bigskip
With this collection of examples and bridges across logic and programming languages, we hope to have convinced the reader that the dictionary we started with continues to receive enrichments, in great part triggered by concepts born in computer science (such as the last mentioned: subtyping).


\begin{thebibliography}{99}
\bibitem{AJ92} S. Abramsky and R. Jagadeesan, New foundations for the geometry of interaction, Information and Computation 111 (1), 53-119 (1994).

\bibitem{AJM}
S. Abramsky, R. Jagadeesan,  and P. Malacaria,
Full abstraction for PCF, Information and Computation 163, 409-470 (2000).
(Manuscript circulated since 1994.)

\bibitem{AMC} S. Abramsky and G. McCusker, Game semantics, in Computational Logic, U. Berger and H. Schwichtenberg eds, Springer-Verlag, 1-56 (1999).

\bibitem{AmaCu} R. Amadio and P.-L. Curien,
Domains and lambda-calculi, Cambridge University Press (1998).

\bibitem{AP-Foc}
J.-M. Andreoli and R. Pareschi, Linear objects: logical processes with built-in inheritance, New Generation Computing 9 (3-4), 445-473 (1991).

\bibitem{BerryCurien82}
G. Berry and P.-L. Curien,
Sequential algorithms on concrete data structures,
Theoretical Computer Science 20, 265-321 (1982).

\bibitem{BerryCurien85}
G. Berry  and P.-L. Curien,
Theory and practice of sequential algorithms: the kernel of the applicative language CDS,
in {\em Algebraic methods in semantics}, Nivat and Reynolds eds,
Cambridge University Press, 35-87 (1985).

\bibitem{Symseq}
P.-L. Curien, On the symmetry of sequentiality,
Proc. 
Mathematical Foundations of Programming Semantics 1993,
Springer Lect. Notes in Comp. Science. 802, 122-130
(1993).

\bibitem{CCF94}
R. Cartwright, P.-L. Curien, and M. Felleisen,
Fully abstract semantics for observably sequential languages,
Information and Computation 111(2), 297-401 (1994).

\bibitem{CH00} P.-L. Curien and H. Herbelin,
The duality of computation, Proc. International Conference on Functional Programming 2000, Montr\'eal, ACM Press (2000).

\bibitem{Coq} T. Coquand, A semantics of evidence for 
classical
arithmetic, Journal of Symbolic Logic 60, 325--337 
(1995).

\bibitem{Fel86} W. Felscher, Dialogues as a foundation 
of intuitionistic logic,
Handbook of Philosophical Logic 3, 341-372 (1986).


\bibitem{Gir87} J.-Y. Girard, Linear logic,
Theoretical Computer Science 50, 1-102 (1987).

\bibitem{Gir89} J.-Y. Girard, Geometry of interaction I: interpretation of system F, in Proc. Logic Colloquium '88, 221-260, North Holland (1989).

\bibitem{Gir01} J.-Y. Girard, Locus Solum, Mathematical Structures in Computer Science (2001).


\bibitem{Grif}
T. Griffin, A formulae-as-types notion of control,
Proc. Principles of Programming Languages 1990, ACM Press (1990).

\bibitem{Her95} H. Herbelin, S\'equents qu'on calcule, Th\`ese de
Doctorat, Universit\'e Paris 7 (1995).

\bibitem{How}
W. Howard, The formulas-as-types notion of construction,
in Curry Festschrift, Hindley and Seldin eds, 479-490 , Academic Press (1980).
(Manuscript circulated since 1969.)
\bibitem{HO}
M. Hyland and L. Ong, On full abstraction for PCF,
Information and Computation 163, 285-408  (2000). (Manuscript circulated since 1994.)

\bibitem{KaMac} G. Kahn, D. Macqueen, Coroutines and networks of parallel processes, in Proc. Information Processing 77, North Holland, 993-998 (1977).

\bibitem{Kleene} S. Kleene, 
Recursive Functionals and Quantifiers of Finite Types Revisited I, II, III, and IV, in Proc. General Recursion Theory II, Fenstad et al. eds, North-Holland (1978), Proc. of the Kleene Symposium, Barwise et al. eds, North-Holland (1980), Proc. Patras Logic Symposium, North Holland (1982), and 
Proc. Symposia in Pure Mathematics 42 (1985), respectively.

\bibitem{Laird} J. Laird, A semantic analysis of control, Ph.D. thesis, University of Edinburgh (1999).

\bibitem{Lam92}
F. Lamarche,
Sequentiality, games and linear logic, manuscript
(1992).

\bibitem{Loa}
R. Loader, Finitary PCF is undecidable, manuscript, University of Oxford (1996).

\bibitem{Mil77}
R. Milner, Fully abstract models of typed lambda-calculi,
 Theoretical Computer Science 4, 1-23 (1977).

\bibitem{Milpi} R. Milner, Communicating and mobile systems: the $\pi$-calculus, Cambridge University Press (1999).

\end{thebibliography}
\end{document}